\documentclass{nature}

\usepackage{graphicx}
\usepackage{textcomp}
\usepackage{amssymb}
\renewenvironment*{figure}{\@float{figure}}{\end@float}
\makeatletter
\let\saved@includegraphics\includegraphics
\AtBeginDocument{\let\includegraphics\saved@includegraphics}
\renewenvironment*{figure}{\@float{figure}}{\end@float}
\makeatother

\usepackage{times}
\usepackage{pgfplots}
\usepackage{geometry}
\usepackage{booktabs}
\usepackage{lineno}
\usepackage{refcount}
\usepackage{xstring}
\newcommand\fref[1]{\IfRefUndefinedExpandable{#1}{??}{S\ref{#1}}}
\newcommand\addrefending[1]{\StrBehind{\getrefnumber{#1}}{.}[\fignumber]%
  \ifcase\fignumber\relax\or de\or se\else te\fi}

\usepackage{color}
\definecolor{myBlue}{rgb} {0,0,1.0} 
\definecolor{myRed}{rgb} {1.0,0,0} 

\title{Coexistence of distinct skyrmion phases observed in hybrid ferromagnetic/ferrimagnetic multilayers}

\author{Andrada-Oana~Mandru$^{1}$, O\u{g}uz~Y{\i}ld{\i}r{\i}m$^{1}$, Riccardo~Tomasello$^{2}$, Paul~ Heistracher$^{3}$, Marcos~Penedo$^{1}$, Anna~Giordano$^{4}$, Dieter~Suess$^3$, Giovanni~Finocchio$^{4}$, and Hans~Josef~Hug$^{1,5}$}

\begin{document}

\maketitle

\begin{affiliations}
 \item Empa, Swiss Federal Laboratories for Materials Science and Technology, CH-8600 D\"{u}bendorf, Switzerland
 \item Institute of Applied and Computational Mathematics, FORTH, GR-70013 Heraklion-Crete, Greece
 \item Christian Doppler Laboratory for Advanced Magnetic Sensing and Materials, Faculty of Physics, University of Vienna, Boltzmanngasse 5, 1090 Vienna, Austria
 \item Department of Mathematical and Computer Sciences, Physical Sciences and Earth Sciences, University of Messina, I-98166 Messina, Italy
\item Department of Physics, University of Basel, CH-4056 Basel, Switzerland
\end{affiliations}

\begin{abstract}

Materials hosting magnetic skyrmions at room temperature could enable new computing architectures as well as compact and energetically efficient magnetic storage such as racetrack memories\,\cite{Zazvorka:2019eo,Fert:2013fq,Tomasello:2014kab,Yu:2017dt,Soumyanarayanan:2016cg,Finocchio:2016dc,Fert:2017bt}. In a racetrack device, information is coded by the presence/absence of magnetic skyrmions forming a chain that is moved through the device. The skyrmion Hall effect that would eventually lead to an annihilation of the skyrmions at the edges of the racetrack can be suppressed for example by anti-ferromagnetically-coupled skyrmions\,\cite{Legrand:2019di}. However, avoiding modifications of the inter-skyrmion distances in the racetrack remains challenging\,\cite{Suess:2018dt}. As a solution to this issue, a chain of bits could also be encoded by two different solitons such as a skyrmion and a chiral bobber\,\cite{Zheng:2018bfc,Ahmed:2018fx}. The major limitation of this approach is that it has solely been realized in B20-type single crystalline material systems that support skyrmions only at low temperatures, thus hindering the efficacy for future applications. Here we demonstrate that a hybrid ferro/ferri/ferromagnetic multilayer system can host two distinct skyrmion phases at room temperature. By matching quantitative magnetic force microscopy data with micromagnetic simulations, we reveal that the two phases represent tubular skyrmions and partial skyrmions (similar to skyrmion bobbers). Furthermore, the tubular skyrmion can be converted into a partial skyrmion. Such multilayer systems may thus serve as a platform for designing skyrmion memory applications using distinct types of skyrmions and potentially for storing information using the vertical dimension in a thin film device.
\end{abstract}

Magnetic skyrmions have been observed in ferromagnets\,\cite{Anonymous:I98Yipv7,MoreauLuchaire:2016em,Woo:2016jwb,Soumyanarayanan:2017gza,Maccariello:2018hka,Li:2019cc}, ferrimagnets\,\cite{Caretta:2018cn,Woo:2018kc}, and synthetic antiferromagnets\,\cite{Legrand:2019di}, and the underlying physics of their stabilization involves a delicate balance of different micromagnetic energies\,\cite{Soumyanarayanan:2016cg,Finocchio:2016dc,Fert:2017bt,Li:2019cc,Legrand:2018gea}. Essentially, the Dzyaloshinskii-Moriya interaction (DMI) breaks the symmetry of the exchange energy and drives the stabilization of chiral structures, such as skyrmions. These are characterized by an integer winding number and hence have peculiar topological properties\,\cite{Nagaosa:2013cc,Fert:2017bt}, which can be used for new device functionalities. Materials hosting two distinct skyrmion phases add an additional degree of freedom in designing devices with improved properties in the emerging field of skyrmionics.  Modeling work showed that such an approach could be used to lower error rates in racetrack memories relying on skyrmions as information carriers\,\cite{Suess:2018dt}. However, to date, the simultaneous existence of two different skyrmion phases, tubular skyrmions and skyrmion bobbers, has only been shown in B20-type FeGe single crystalline materials\,\cite{Zheng:2018bfc}, and in MBE-grown epitaxial FeGe on Si(111) substrates,\cite{Ahmed:2018fx}. Although they show very interesting physics, such single crystalline systems restrict the choice of materials for additional layers needed in devices for example to provide spin-orbit torque, an exchange field or simply for electrons used for applying currents or measuring the skyrmion induced Hall signals\,\cite{Zhang:2020ce}.

Here, we develop a thin film system by combining two ferromagnetic (SK) layers separated by a ferrimagnetic (Fi) layer with interfacial DMI, designed to support two distinct skyrmion phases at room temperature that is in principle compatible with other layers needed for device functionality. Our quantitative magnetic force microscopy (MFM) analyses performed on a series of samples designed for disentangling the MFM contrast arising from different layers clearly show the stabilization of a tubular skyrmion, which is extended in all the ferromagnetic and ferrimagnetic layers, and an incomplete skyrmion nucleated in the ferromagnetic layers only. The diameters of the tubular and incomplete skyrmions are noticeably different, thus making these configurations suitable for a direct electrical detection by measuring the magnetoresistive signal. The experimental findings are supported quantitatively by micromagnetic simulations, demonstrating that the key ingredient to achieve the coexistence of the two skyrmion phases is the design of a Fi layer with a sufficiently large DMI. This work paves the way for the development of hybrid systems by merging together a variety of properties from different materials to move the skyrmionics a step forward towards practical applications. In particular, this work can impact the design of new robust skyrmion-based memory and computing architectures working at room temperature and coding the information in two types of skyrmions.

Our sample contains two 14.5\,nm thick [Ir(1)/Fe(0.3)/Co(0.6)/Pt(1)]$_{\times 5}$-multilayers (SK layers) separated by a 3.8\,nm thick Fi layer (SK/Fi/SK, see Fig.\,1{\bf a}). 
\begin{figure}[t]
\renewcommand{\baselinestretch}{1} 
    \centering
    \includegraphics[width=150mm]{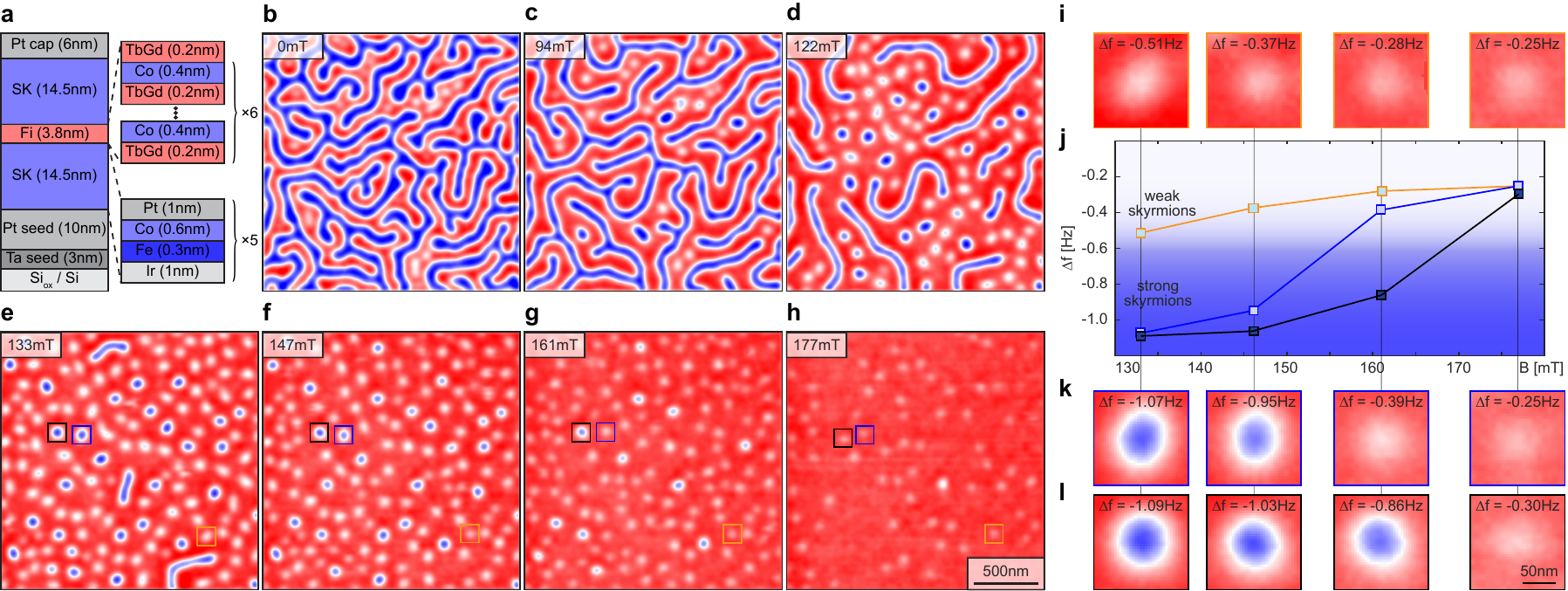}
    \caption{\small MFM results obtained on a trilayer sample supporting two different skyrmion phases. {\bf a} Schematics of the trilayer sample consisting of ferromagnetic top and bottom layers (SK), and a ferrimagnetic layer (Fi). These three main layers again consists of a multilayer to achieve specific magnetic properties. {\bf b}-{\bf h} MFM data acquired in external fields $\mu_0H_z =0-177$\,mT. Panels {\bf i}, {\bf k}, and {\bf l} show magnified views of the skyrmions highlighted by the black, blue and orange squares in panels {\bf e}-{\bf h} for fields of 133, 147, 161, and 177\,mT. Panel {\bf j} displays the evolution of the skyrmion contrast with the applied field.  }
\end{figure}
Previous studies\,\cite{Soumyanarayanan:2017gza} demonstrated that the Ir/Fe\,\cite{Heinze:2011ica} and Co/Pt\,\cite{Yang:2015hc} interfaces in the Ir/Fe/Co/Pt-multilayers exhibit a strong negative DMI supporting skyrmions with a clockwise N{\'e}el wall texture at room temperature. The composition and layer thickness of the Fi layer permit an independent adjustment of its anisotropy and magnetization by the Tb:Gd ratio and rare-earth to Co-layer thickness, respectively. Here, a 1:1 Tb:Gd ratio was selected to obtain a weak 
effective perpendicular anisotropy $K_{\rm eff} =336 \pm 23\, $kJ/m$^3$, determined by magnetometry measurements (see Fig.\,S1 of Supplementary Note 1). Such a weak positive anisotropy supports perpendicular magnetization structures with domains having a size up to several tens of microns.
The SK/Fi/SK sample grown here is designed in such a way that skyrmions generated by the top and bottom SK layers could enter the Fi layer stabilizing skyrmion-like tubular structures penetrating through all layers of the sample. Consequently, high-resolution MFM has been used to study the micromagnetic state of the sample and its evolution as a function of the applied field. 

Figures\,1{\bf b}-{\bf h} show MFM frequency shift  ($\Delta f$) data with a total range of 1.4\,Hz acquired at room temperature in fields $\mu_0 H_z$ ranging from 0 to 177\,mT. 
MFM data of the remanent state show a red/blue (up/down) maze domain pattern with about $\pm 0.7\,$Hz of $\Delta f$-contrast (Fig.\,1{\bf b}). If the field is increased from zero to $\mu_0 H_z = 122$\,mT (Fig.\,1{\bf d}), the red (up) domains expand and inside these, skyrmions become more pronounced. At a field of $\mu_0 H_z = 133$\,mT (Fig.\,1{\bf e}), almost all stripe domains collapse and skyrmions generating two distinct MFM signal levels (appearing with dark blue and white color) are observed. If the field (applied antiparallel to the skyrmion cores) is increased from 133-177\,mT (Fig\,1{\bf e}-{\bf h}), the MFM contrast of all skyrmions is reduced and some of them are annihilated. However, while the MFM signal of the weak (white) skyrmions becomes gradually smaller with increasing field, the contrast of the strong (dark blue) skyrmions drops to the level of the weak (white) skyrmions at some critical field value (161 or 177\,mT) after an initial gradual contrast decay of their MFM signal at lower fields (Figs.\,1{\bf i}-{\bf l}). 

The MFM contrast generated by a skyrmion arises from the convolution of its stray field above the sample surface with the magnetic charge distribution of the MFM tip\,\cite{Bacani:2019bm}. Consequently, the radius of a skyrmion in an MFM image is wider than that of its spin texture. A reduction of the MFM signal in increasing fields is therefore compatible with a reduction of the skyrmion radius (see Fig.\,S2 of Supplementary Note 2). The distinct drop of the MFM signal observed for the strong skyrmions at fields of 161 or 177\,mT can however not be explained by a gradual field-driven reduction of the skyrmion radius, but is indicative of a field-induced switching of the skyrmion type occuring in the SK/Fi/SK sample.

For such a seemingly complicated sample hosting two types of skyrmions, it is important to understand the contributions coming from different layers.
In order to disentangle the MFM contrasts, a series of additional samples have been fabricated: three consisting of selected parts of the SK/Fi/SK sample, and one in which the Fi layer was replaced by a magnetically inactive Ta layer of the same thickness. These sample structures are depicted in Figs.\,2{\bf a}-{\bf e} together with their corresponding MFM data (Figs.\,2{\bf f}-{\bf j}), all acquired in a field of about 133\,mT. Also visible are zoomed views of selected skyrmions, fitted Gaussian functions, and the fitted skyrmion contrast values (Figs.\,2{\bf k}-{\bf o}). 
Note that in order to quantitatively compare the MFM signals obtained on different samples, a recently developed frequency modulated distance feedback method\,\cite{Zhao:2018kd} was used. It 
permits keeping the tip-sample distance constant with a precision of about 0.5\,nm over many days, even after re-approaching the tip on different samples and in applied magnetic fields. This can be achieved without ever bringing the tip in contact with the sample surface such that the magnetic coating of the tip remains intact and the same tip can be used for all samples.

\begin{figure} 
\renewcommand{\baselinestretch}{1} 
    \centering
    \includegraphics[width=150mm]{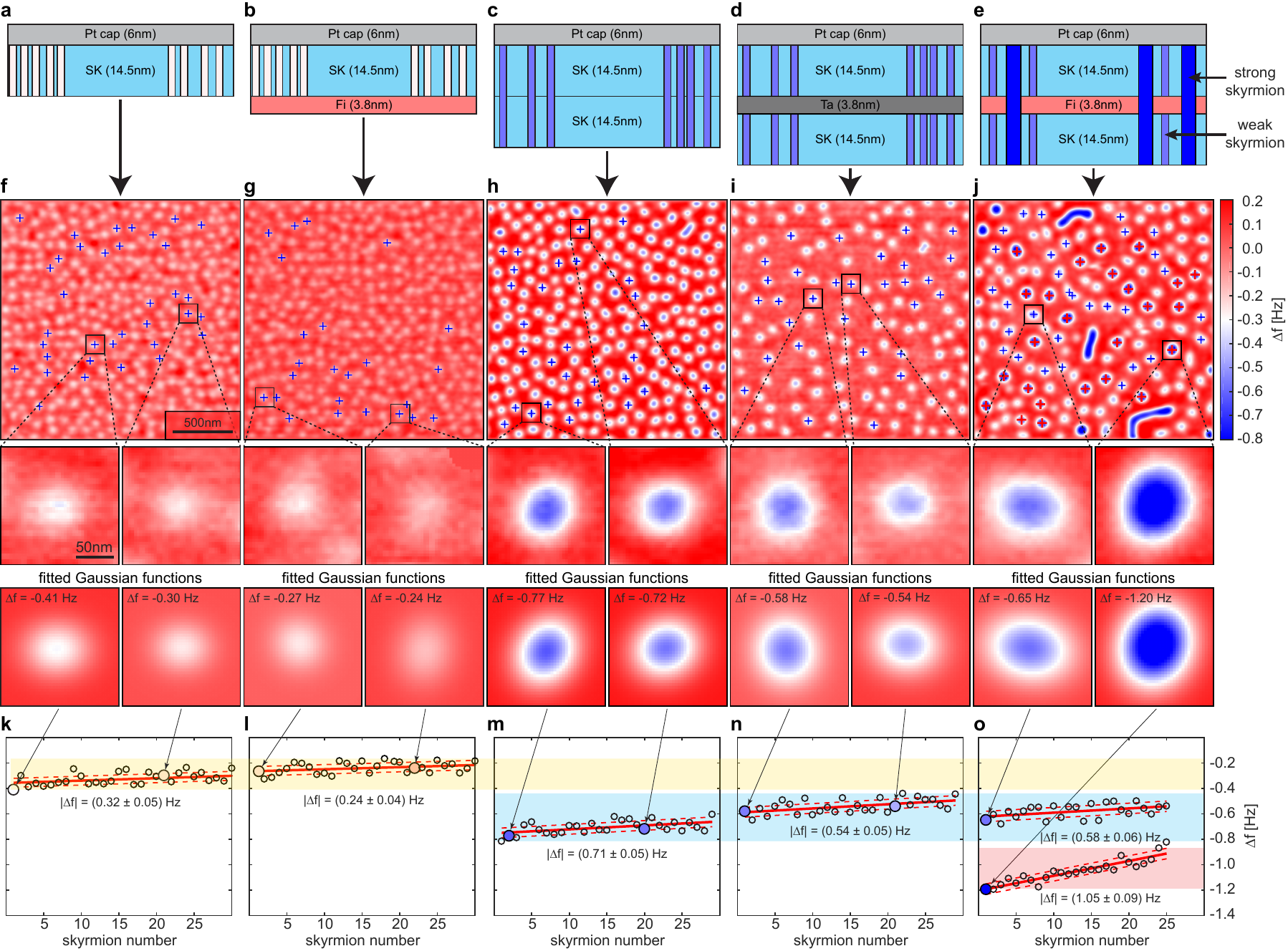}
    \caption{\small Schematics of different samples used to disentangle the contributions to the measured MFM contrast on the SK/Fi/SK sample. All images have been taken in a field of about 133\,mT. {\bf a} A single 14.5\,nm thick ferromagnetic layer (SK) is deposited directly on the Pt-seed (not shown) on top of the oxidized Si-wafer. The vertical, white rectangles schematically depict the skyrmions. {\bf b} The same ferromagnetic layer as in  {\bf a} but deposited on the [(TbGd)(0.2)/Co(0.4)]$_{\times 6}$/(TbGd)(0.2) ferrimagnetic (Fi) layer. {\bf c} A sample consisting of two 14.5\,nm thick SK layers. {\bf d} Similar to {\bf c} but the two SK layers are separated by a magnetically inactive 3.8\,nm thick Ta layer. The skyrmions within the two SK layers are depicted by the vertical pale blue rectangles. {\bf e} Schematics of the trilayer sample consisting of two SK layers separated by a 3.8\,nm thick Fi layer; the skyrmions are depicted by the vertical, white and pale blue rectangles penetrating through the SK layer(s), and by the wider dark blue rectangles for the strong contrast skyrmions penetrating through all layers. {\bf f}-{\bf j} $2\mu$m$\times 2\mu$m  MFM data (top row) and zoomed images of skyrmions shown together with the fitted Gaussian functions (lower rows) obtained on each of the samples in {\bf a}-{\bf e}. The crosses in panels {\bf f} to {\bf j} highlight the skyrmions which have been fitted to determine the $\Delta f$-contrast displayed in panels {\bf k} to {\bf o}. } 
\end{figure}

For the sample consisting of a single SK layer ([Ir(1)/Fe(0.3)/Co(0.6)/Pt(1)]$_{\times 5}$ grown directly on Pt (Fig.\,2{\bf a}), the MFM data (Fig.\,2{\bf f}) reveals a disordered pattern of skyrmions with an areal density comparable to that observed in previous work\,\cite{Soumyanarayanan:2017gza} for similar samples with 20 Ir/Fe/Co/Pt repetitions. This confirms the existence of a strong negative DMI arising from the Ir/Fe\,\cite{Heinze:2011ica} and Co/Pt\,\cite{Yang:2015hc} interfaces. The observed small variation of the MFM skyrmion contrast is attributed to variations of the local values of the DMI, perpendicular anisotropy and exchange stiffness\,\cite{Bacani:2019bm}. The average skyrmion contrast $|\Delta f | = (0.32\pm 0.05)\,$Hz is obtained by fitting selected skyrmions (marked by the blue crosses) with 2D Gaussian functions. Zoomed MFM images of the skyrmions marked by the black squares, the fitted Gaussian functions, and the peak $\Delta f$ signal obtained from the fit are shown below Fig.\,1{\bf f} and in Fig.\,2{\bf k}. If the SK layer is grown directly onto the Fi layer (Fig.\,2{\bf b}), the 
$|\Delta f |$-contrast is reduced to $(0.24 \pm 0.02)$\,Hz (Figs.\,2{\bf g} and {\bf l}), but the areal density of the skyrmions remains about the same. Figure\,2{\bf h} shows MFM data acquired on a [Ir(1)/Fe(0.3)/Co(0.6)/Pt(1)]$_{\times 10}$-multilayer sample (Fig.\,2{\bf c}), representing two SK layers on top of each other. The observed skyrmion $|\Delta f |$-contrast in this case is $(0.71 \pm 0.05)$\,Hz (Fig.\,2{\bf m}), slightly more than double the contrast observed for a sample consisting of a single SK layer (Fig.\,2{\bf f}). If the two SK layers are separated by a  3.8\,nm thick Ta layer (Fig.\,2{\bf d}), the skyrmion $|\Delta f |$-contrast drops to $0.54\pm0.03\,$Hz (Figs.\,2 {\bf i} and {\bf n}). Note that this contrast level agrees well to the $|\Delta f | = (0.58\pm 0.06)\,$Hz generated by the weak skyrmions in the SK/Fi/SK sample (Figs.\,2{\bf e}, {\bf j}, and {\bf o}), where the two SK layers are separated by the Fi layer. However, the contrast remains much weaker than the $|\Delta f | = (1.05\pm 0.09)\,$Hz observed for the strong skyrmions in the same sample. These observations suggest that the weak contrast arises from skyrmions existing solely in the bottom and top SK layers, whereas the strong skyrmion contrast is caused by a tubular skyrmion running through all three layers (see rectangles in Fig.\,2{\bf e} schematically representing the skyrmions). The much stronger contrast of the tubular skyrmions indicates that these spin textures must have a wider radius because their slightly increased length would not lead to the observed pronounced contrast increase.

\begin{figure}[ht]
\renewcommand{\baselinestretch}{1} 
    \centering
    \includegraphics[width=150mm]{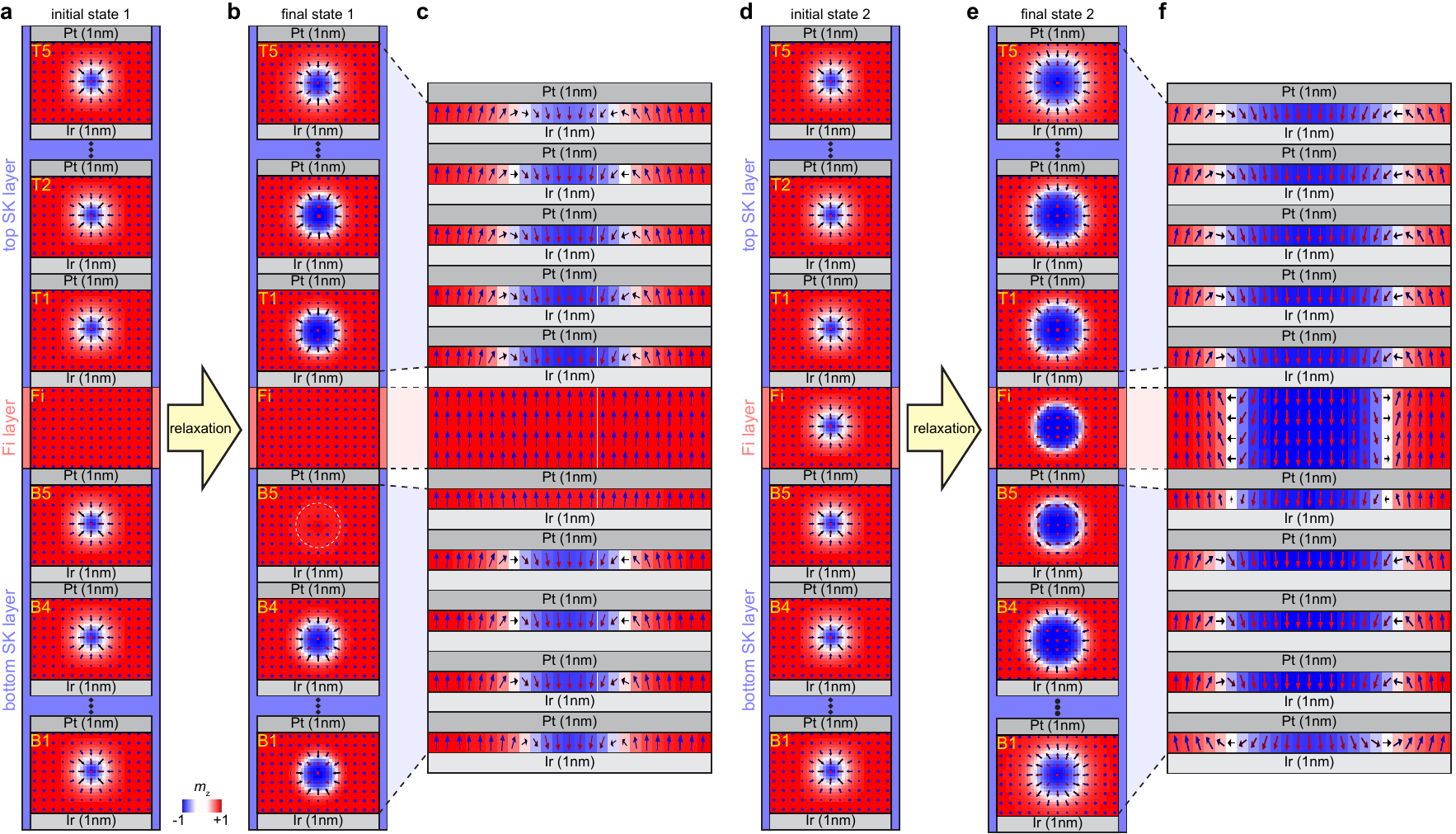}
    \caption{\small {\bf a} Intial state 1 with a clockwise skyrmion spin texture imposed in the bottom and top SK layers but not in the Fi layer. {\bf b} and {\bf c} final state 1 and its corresponding cross-section after relaxation of the initial state 1. A skyrmion spin texture exists solely in the sublayers B1 to B4 of the bottom SK layer and in all sublayers of the top SK layer, but not in the Fi layer. {\bf d} Initial state 2 with a skyrmion spin texture enforced in all layers. {\bf e} and {\bf f}  final state 2 and its corresponding cross-section of: a tubular skyrmion running through all layers is stabilized. 
}
\end{figure}

The physics behind the experimental stabilization of the different skyrmion spin textures is captured by 3D micromagnetic calculations and the results are summarized in Fig.\,3 (see Methods and Table S1 in Supplementary Note 1 for the parameters used). Figure\,3{\bf a} shows an initial state with a clockwise skyrmion spin texture in the bottom and top SK layers and a uniform up magnetization state in the Fi layer (initial state 1). After a relaxation process and for a DMI of the Fi layer, $D_{\rm Fi} = +0.8\,$mJ/m$^2$, a clockwise skyrmion spin texture with a slightly larger radius appears in the four bottom-most Fe/Co-sublayers, B1 to B4 of the bottom SK layer, as well as in all Fe/Co-sublayers, T1 to T5 of the top SK layer (final state 1, see Fig.\,3{\bf b}). As also visible from the cross-section in Fig.\,3{\bf c}, the skyrmion diameter is thickness dependent, being larger near the middle of the sample and smaller in the external layers (Fig.\,3{\bf c}), as expected from the minimization of the magnetostatic energy, and already observed for other simpler structures\,\cite{Legrand:2018gea,Li:2019cc} (see Fig.\,S3{\bf a} in Supplementary Note 3 for the dependence of the skyrmion radius on the layer position). The chirality is the same in all the layers, as expected from the energy minimization when the DMI is large enough with respect to the magnetostatic energy\,\cite{Legrand:2018gea,Li:2019cc} (the DMI of the SK layer is $-2.5\,$mJ/m$^2$). The magnetization of the Fi layer remains uniform and the ferromagnetic interlayer exchange coupling (IEC) through the 1\,nm thick Pt also suppresses the initial skyrmion in the top-most Fe/Co-sublayer, B5, of the bottom SK layer. Therefore, an incomplete skyrmion is obtained.

If a clockwise initial skyrmion spin texture is also imposed within the Fi layer (initial state 2, displayed in Fig.\,3{\bf d}), more complex spin textures develop provided $|D_{\rm Fi}| > 0.7\,$mJ/m$^2$ (see also Fig.\,S3{\bf b} in Supplementary Note 3). The final state 2 and its corresponding cross-sectional view are shown in Figs.\,3{\bf e} and {\bf f}, respectively, for $D_{\rm Fi} = 0.8\,$mJ/m$^2$.
Differently from final state 1, skyrmions exist in all sublayers of the bottom and top SK layers, and also in the Fi layer. Hence, a tubular skyrmion with a larger radius is stabilized. Interestingly, the skyrmion chirality is thickness-dependent. In the bottom-most Fe/Co-sublayer of the bottom SK layer, a N{\'e}el skyrmion with a counter-clock wise chirality opposite to the one favored by a negative DMI is stabilized. This chirality leads to an improved magnetic flux closure and hence optimizes the magnetostatic energy. Layers B2 up to B4 have the clockwise chirality expected for the negative DMI of the SK layers\,\cite{Heinze:2011ica}, whereas the skyrmion chiralities in B5 and Fi layers derive from the trade-off among negative DMI of the SK layer, positive $D_{\rm Fi}$ and IEC. The skyrmion chirality in the Fi layer is intermediate between N{\'e}el outward and Bloch types, thus reminiscent to a counter-clockwise skyrmion expected for the positive $D_{\rm Fi}$\,\cite{Yang:2015hc}. The frustration arising from the negative DMI in the SK layer, positive $D_{\rm Fi}$ and IEC then explains the Bloch-type skyrmion obtained in layer B5, that compromises between clockwise and counter-clockwise chiralities. Furthermore, this frustration also decreases slightly the skyrmion diameter in both the Fi and B5 layers (see Fig.\,S4 of Supplementary Note 3 for $D_{\rm Fi}>0$).

In summary, the hybrid ferro/ferri/ferromagnetic multilayer system presented here supports the coexistence of two skyrmion phases at room temperature: a smaller-diameter incomplete skyrmions existing solely in the top and bottom ferromagnetic layers, and larger-diameter tubular skyrmions running through the entire sample. In future devices this may facilitate the electrical detection of the skyrmions and the distinction between the two states. Moreover, our metallic multilayer system permits the future implementation of additional layers, e.g. to generate a strong-spin orbit torque, or layers providing an RKKY-type exchange to a layer with a perpendicular magnetization to allow the existence of skyrmions at zero field. The concept discussed here thus paves the way for future skyrmionic devices potentially including systems permitting the storage of information along the third dimension.

\section{Methods}
\subsection{Magnetic Force Microscopy}
The MFM measurements were performed using a home-built high-vacuum ($\approx$ 10$^{-6}$\,mbar) MFM system equipped with an {\it in-situ} magnetic field of up to $\approx$ 300\,mT. By operating the MFM in vacuum, we obtain a mechanical quality factor $Q$ for the cantilever of $\approx$ 200,000. Using such high $Q$ values improves the sensitivity by a factor of about 40 compared to MFM performed in air and also permits the use of a thin magnetic coating on the tip. SS-ISC cantilevers from Team Nanotech GmbH with a tip radius below 5\,nm (without any initial coating) were used. In order to make the cantilever tip sensitive to magnetic fields, we sputter-deposited at room temperature 3\,nm of Co on a Ta seed (2\,nm) and then capped with Ta (4\,nm) to prevent oxidation. A Zurich Instruments phase locked loop (PLL) system was used to oscillate the cantilever on resonance at a constant amplitude of 7\,nm and to measure the frequency shift arising from the tip-sample interaction force derivative. Note that the frequency shift is negative for an attractive force (derivative). For the MFM data shown in Figs.\,1 and 2, an up field was applied and an MFM tip with an up magnetization was used. Therefore, the skyrmions have a down magnetization as those in our micromagnetic simulations (Fig.\,3). The up tip magnetization and the down magnetization of the skyrmions then generates a positive frequency shift contrast that would correspond to a red color in the MFM images displayed in Figs.\,1 and 2. In order to facilitate the comparison of the skyrmions measured by MFM with those obtained from micromagnetic calculations, the MFM data from Figs.\,1 and 2 have been inverted. 

\subsection{Sample Preparation}
Samples were grown using DC magnetron sputtering under a 2\,{\textmu}bar Ar atmosphere using an an AJA Orion system with base pressure of $\approx$ 1$\times$10$^{-9}$\,mbar. All multilayers were deposited onto thermally oxidized Si(100) substrates with Ta(3\,nm)/ Pt(10\,nm) as seed layers and Pt(6\,nm) as capping layer (for oxidation protection). The substrates were annealed at $\approx$ 100$\,^{\circ}$C for an hour and cooled down close to room temperature before each deposition. The layer thickness was determined by calibrations performed using X-ray reflectivity on samples containing single layers of each individual element. Since the single SK layer sample is very sensitive to the Fe and Co thicknesses, the reproducibility was verified periodically by re-growing such a sample and performing MFM measurements under the same conditions.

\subsection{Magnetometry Measurements}
The bulk magnetic properties of the samples were determined by vibrating sample magnetometry (VSM) using a 7\,T Quantum Design system. The measurements were performed at 300\,K for both {\it in-plane} and {\it out-of-plane} geometries and in fields of up to 4\,T. All samples were measured using the same VSM holder and each measurement was repeated several times. In addition, the background signal coming from the VSM holder and bare substrate was periodically checked to ensure a clean magnetic signal coming from the ferro- and/or ferrimagnetic layers only.

\subsection{Micromagnetic Simulations}
The micromagnetic computations were carried out by means of a state-of-the-art micromagnetic solver, PETASPIN\,\cite{Giordano:2012gw} and magnum.af
\,\cite{heistracher2020hybrid}, both based on the finite difference scheme and which numerically integrate the Landau-Lifshitz-Gilbert (LLG) equation by applying the Adams-Bashforth\, time solver scheme:   
\begin{equation}
\frac{d{\bf m}}{d\tau} = -({\bf m} \times {\bf h}_{\rm eff}) + \alpha _{\rm G} \left( {\bf m} \times \frac{d{\bf m}}{d\tau} \right) \,,
\label{Giovannis eq}
\end{equation}
where $\alpha _{\rm G}$ is the Gilbert damping, ${\bf m } = {\bf M} / M_{\rm s}$ is the normalized magnetization, and $\tau = \gamma _0 M_{\rm s} t$ is the dimensionless time, with $\gamma _0$ being the gyromagnetic ratio, and $M_{\rm s}$ the saturation magnetization. $ {\bf h}_{\rm eff} $ is the normalized effective field in units of $M_{\rm s}$, which includes the exchange, interfacial DMI, magnetostatic, anisotropy and external fields\,\cite{Li:2019cc,Tomasello:2014ix}. The DMI is implemented as
\begin{equation}
\epsilon_{\rm InterDMI} = D\left[ m_z \nabla \cdot {\bf m} - ({\bf m} \cdot \nabla) m_z \right]
\label{Dieters eq}
\end{equation}

The [Ir(1\,nm)/Fe(0.3\,nm)/Co(0.6\,nm)/Pt(1\,nm)]$_5$ SK layers are simulated by 5 repetitions of a 1\,nm thick CoFe ferromagnet separated by a 2\,nm thick Ir/Pt non-magnetic layer. Each ferromagnetic layer is coupled to the other ones by means of the magnetostatic field only (exchange decoupled); for simplicity, we neglect any Ruderman-Kittel-Kasuya-Yosida (RKKY) interactions. For the SK layer, we used the following physical parameters: saturation magnetization $M_{\rm s} = 1371 \pm 41\, $kA/m, and uniaxial perpendicular anisotropy constant $K_{\rm u} =1316 \pm 92\, $kJ/m$^3$ (both obtained by our VSM measurements), exchange constant $A=15\,$pJ/m, and 
interfacial DMI constant $D = -2.5\,$mJ/m$^2$ from\,\cite{Soumyanarayanan:2017gza}. The ferrimagnetic [(TbGd)(0.2\,nm)/Co(0.4\,nm)]$_6$/(TbGd)(0.2\,nm)-multilayer s simulated by a 4\,nm magnetic layer. Its saturation magnetization $M_{\rm s,Fi} = 488 \pm 34\, $kA/m, equal to the net magnetization of the experimental ferrimagnet, and its uniaxial perpendicular anisotropy constant $K_{\rm u,Fi} = 486 \pm 44\, $kJ/m$^3$ were again measured by VSM. The exchange constant $A_{\rm Fi} = 4\,$pJ/m was used in agreement with our prior work for rare-earth-transition metal alloy layers\,\cite{Zhao:2019gn}. We use a 
discretization cell size of $3 \times 3 \times 1\,$nm$^3$.
The top ferromagnetic layer (B5) of the bottom SK layer is coupled to the first 1\,nm of the ferrimagnetic layer via an RKKY-like interlayer exchange coupling\,\cite{Tomasello:2017hy}. We set a positive value of the constant (ferromagnetic coupling) equal to $0.8\,$mJ/m$^2$ from\,\cite{Omelchenko:2018gia}. In all the simulations, an out-of-plane external field $H_{\rm ext} = 130\,$mT is applied antiparallel to the skyrmion core.\\



\section*{References}


\begin{addendum}
 \item[Acknowledgements] A.-O.M., O.Y. and H.J.H. thank the Swiss National Science Foundation under Projects 200021-147084,
 200021E-160637, 154410, and Empa for the financial support. R.T. and G.F. thank the project ThunderSKY funded from the Hellenic Foundation for Research and Innovation (HFRI) and the General Secretariat for Research and Technology (GSRT) under Grant No. 871 and the support by PETASPIN association. D.S. acknowledges the Austrian Science Fund under Grant I2214-N20 for financial support.

 \item[Author Contributions] H.J.H. and G.F. conceived the idea and planned the multidisciplinary approach to develop the project. H.J.H., A.-O.M and O.Y. designed the multilayer systems. O.Y. and A.-O.M grew the samples based on earlier samples grown and measured by M.P.; A.-O.M. performed the MFM experiments. O.Y. carried out the VSM measurements. H.J.H. performed the MFM data analysis. R.T. performed the micromagnetic simulations. A.G. developed the software to simulate hybrid ferromagnetic/ferrimagnetic multilayer systems. R.T. and G.F. analyzed and interpreted the micromagnetic data. D.S. and P.H. performed micromagnetic calculations to obtain the stray field above the sample. All authors contributed to the writing of the manuscript.

 \item[Competing Interests] The authors declare that they have no
competing financial interests.

 \item[Correspondence] Correspondence and requests for materials
should be addressed to A.-O.~Mandru~(email: andrada-oana.mandru@empa.ch).
\end{addendum}


\end{document}


\maketitle

\begin{affiliations}
 \item Empa, Swiss Federal Laboratories for Materials Science and Technology, CH-8600 D\"{u}bendorf, Switzerland
 \item Institute of Applied and Computational Mathematics, FORTH, GR-70013 Heraklion-Crete, Greece
 \item Christian Doppler Laboratory for Advanced Magnetic Sensing and Materials, Faculty of Physics, University of Vienna, Boltzmanngasse 5, 1090 Vienna, Austria
  \item Department of Mathematical and Computer Sciences, Physical Sciences and Earth Sciences, University of Messina, I-98166 Messina, Italy
\item Department of Physics, University of Basel, CH-4056 Basel, Switzerland
\end{affiliations}

\pagebreak

 \section*{ \centering Supplementary note 1. Magnetometry measurements and material parameters used in the micromagnetic simulations}

 \begin{figure}[ht]
\centering
\includegraphics[width=150mm]{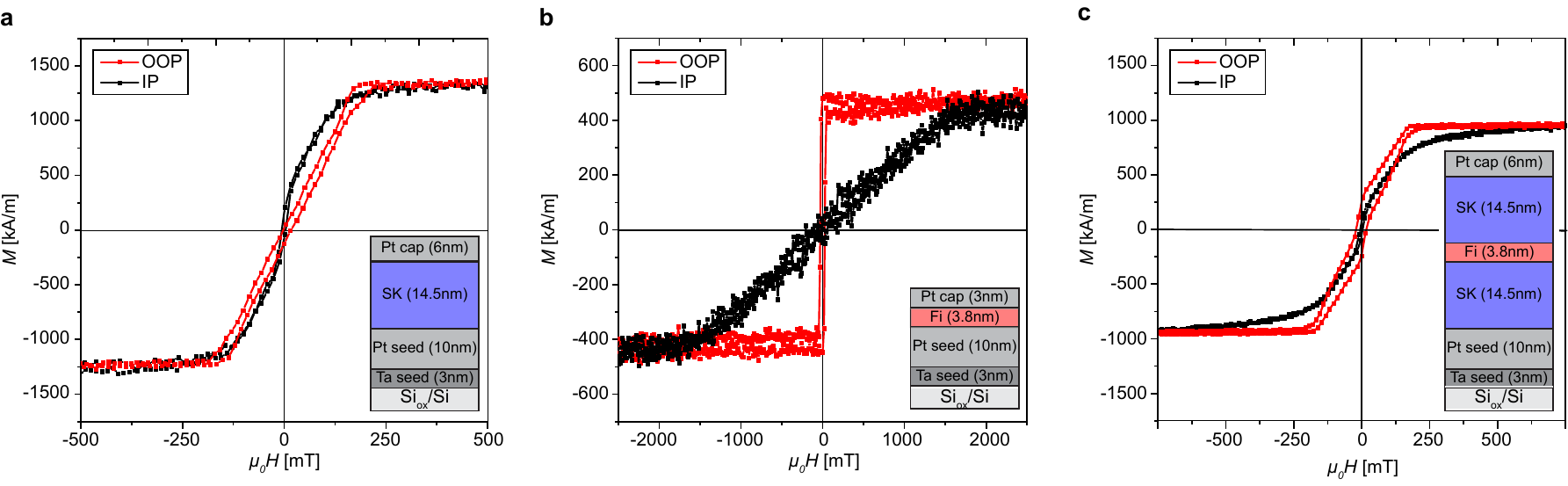}
\caption{\small {\it In-plane} (IP, black) and {\it out-of-plane} (OOP, red) M-H loops acquired at 300 K on \textbf{a} the single ferromagnetic SK layer, \textbf{b} the Fi layer, and \textbf{c} the trilayer samples with corresponding layer schematics. }
\label{fig:S1}
\end{figure}
 The magnetometry measurements for the single SK layer, the Fi layer, and trilayer samples are presented in Fig.\,\fref{fig:S1}{\bf a}-{\bf c}. Note that all measurements were performed with a magnetic field range of $\pm$ 4\,T, well above the saturation field of all samples. However, for clarity purposes, we show the zoomed-in data. The backgrounds coming from the {\it in-plane} and {\it out-of-plane} holders (together with a Si substrate) were measured independently and then subtracted from all loops. The magnetization values were calculated by considering only the ferromagnetic layers: for the SK layer sample the total thickness $t_{\rm Fe}+t_{\rm Co} = 4.5$\,nm was considered, and for the trilayer sample the additional $t_{\rm Fe}+t_{\rm Co} = 4.5$\,nm and the Fi layer thickness of 3.8\,nm were taken into account. From these measurements, the ani-sotropy field $H_{\rm a}$, saturation magnetization $M_{\rm s}$, and the uniaxial perpendicular anisotropy constant $K_{\rm u}$ for the single SK and Fi layers were extracted and served as part of the input for the micromagnetic simulations. 
 \begin{table}
 	\centering
 	\vspace{0.5cm}
 	\begin{tabular}{l c c}
 		\toprule
 		\toprule
 		layer & SK &  Fi\\
 		\midrule
 		$M_{\rm s}$\,[kA/m] & 1371$^*$ & 488\\
 		$K_{\rm u}$\,[kJ/m$^3$ ] &  1316$^*$ & 486\\
 		$A_{\mathrm{ex}}$\,[pJ/m] & 15 & 4 \\        
 		$D $\,[mJ/m$^2$ ] & -2.5 & +0.8 \\
 		$RKKY_{\rm Pt}$\,pJ/m & 4 \\
 		\bottomrule
 		\bottomrule
 	\end{tabular}
 	\caption{ {\small Material parameters of the SK ([Ir(1)/Fe(0.3)/Co(0.6)/Pt(1)]$_{\times 5}$) and Fi ([(TbGd)(0.2)/Co(0.4)]$_{\times 6}$/(TbGd)(0.2)) layers used in the micromagnetic simulations. $^*$Note that for the calculation of the magnetization  $M_{\rm s}$ and the anistropy $K_{\rm u}$ of the SK layer all magnetic moments are attributed to the Fe(0.3)/Co(0.6)-layers.}}
 	\label{tab:mat2}
 \end{table}

\pagebreak
\section*{ \centering Supplementary note 2: Skyrmion stray field calculations}
 \begin{figure}[hb]
\centering
\includegraphics{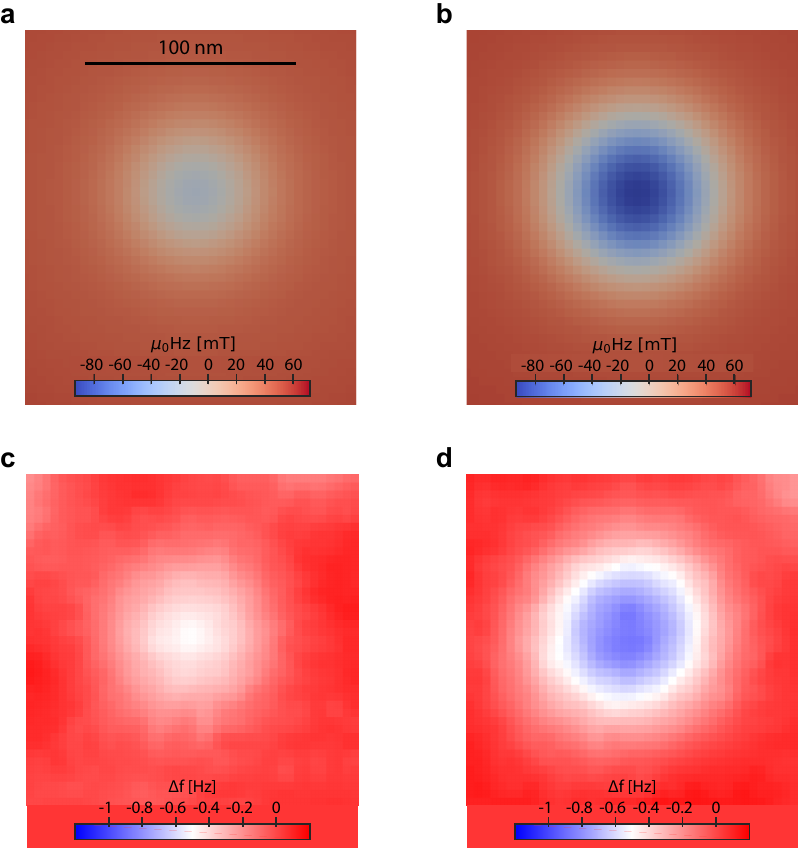}
\caption{\small Z-component of the stray field generated by the incomplete {\bf a} and tubular {\bf b} skyrmion. {\bf c} and {\bf d} MFM $\Delta f$-images of an incomplete and a tubular skyrmion, respectively. These particular skyrmions show about the same average contrast of the weak (0.58\,Hz) and strong (1.05\,Hz) contrast skyrmions visible in Fig.\,2{\bf j} of the main manuscript.}
\label{fig:S2}
\end{figure}
Figure\,\fref{fig:S2} depicts the z-component of the stray field generated by each of the two stable skyrmion states, where the weak skyrmion refers to state 1 in Fig.\,3{\bf b} and the strong skyrmion refers to state 2 in Fig.\,3{\bf e} of the main manuscript. The field is obtained from micromagnetic simulations and evaluated at a distance of 21\,nm above the top-most ferromagnetic layer, in accordance with the 6\,nm-thick capping Pt layer and the tip-sample distance of about 15\,nm used all MFM measurements shown in this study.
As seen from Fig.\,\fref{fig:S2}{\bf a}, the H$_z$ field varies between -22.7\,mT and 70.9\,mT; from Fig.\,\fref{fig:S2}{\bf b} we obtain a range between -93.3\,mT and 71.4\,mT. The 1.76 times larger field difference for the case of the tubular skyrmion agrees well with the 1.81 times higher MFM contrast obtained when comparing the strong skyrmions to the weak ones. 

\pagebreak
 \section*{ \centering Supplementary note 3:\\  
 Skyrmion diameter and chirality as a function of layer position and $D_{\rm Fi}$}
\begin{figure}[ht]
\centering
\includegraphics{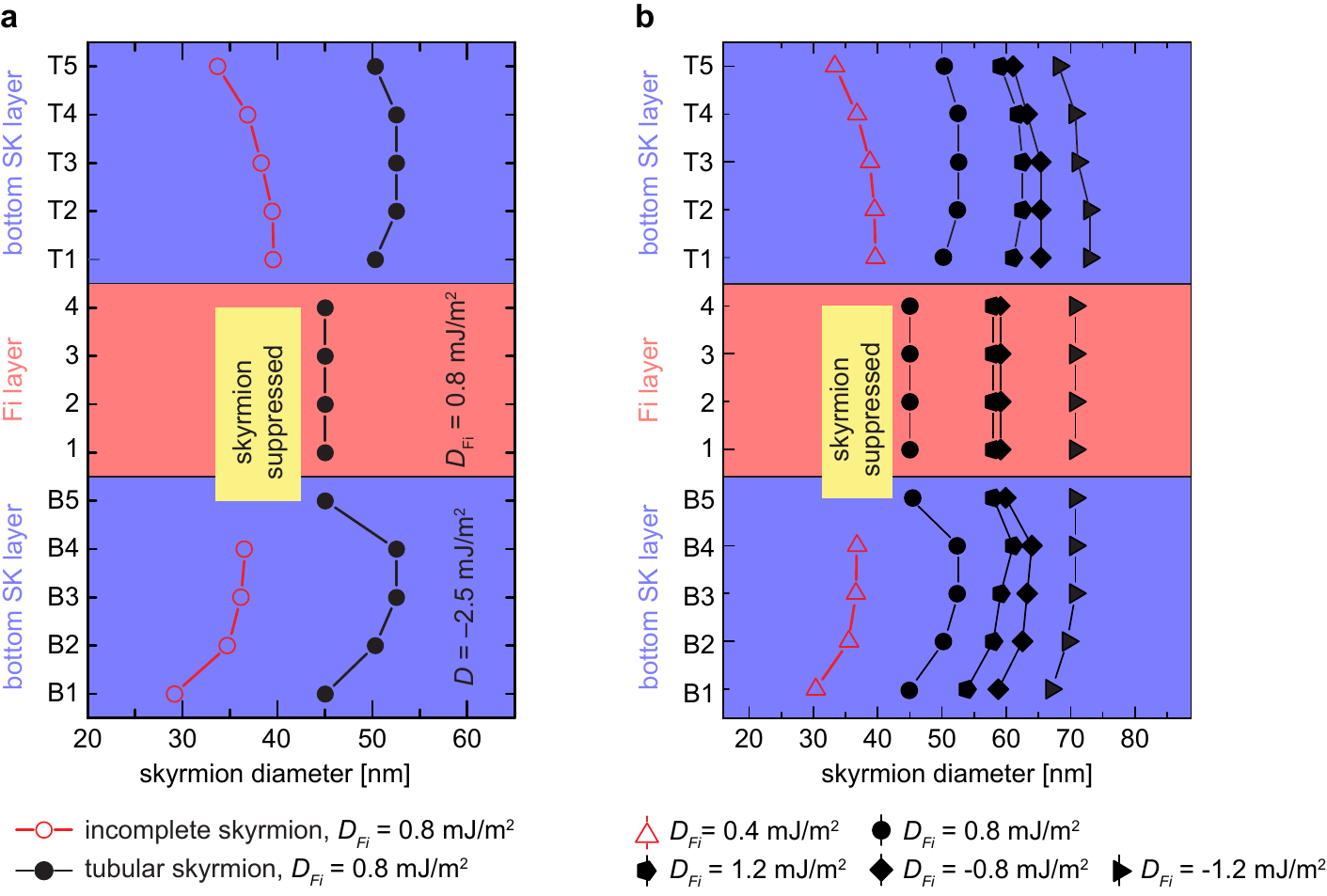}
\caption{\small {\bf a} Skyrmion diameter as a function of the layer position for the incomplete skyrmion and tubular skyrmion. The DMI of the SK layers is $D = -2.5\,$mJ/m$^2$, while that of the interlayer is $D_{\rm Fi} = 0.8\,$mJ/m$^2$.
{\bf b} Skyrmion diameter as a function of layer position and $D_{\rm Fi} = 0.4\,$mJ/m$^2$ (red, up, open triangles and lines), $0.8\,$mJ/m$^2$ (circles), $1.2\,$mJ/m$^2$ (pentagons), $-0.8\,$mJ/m$^2$ (diamonds), and $-1.2\,$mJ/m$^2$ (right triangles).}
\label{fig:S3}
\end{figure}

The skyrmion diameter as a function of layer position and for different values of $D_{\rm Fi}$ are presented in Fig.\,\fref{fig:S3}. More specifically, Fig.\,\fref{fig:S3}{\bf a} shows the results linked to Fig.\,3 of the main text, i.e. $D_{\rm Fi} = 0.8\,$mJ/m$^2$ for an incomplete (open red circles) and a tubular skyrmion (filled black circles). The skyrmion diameters obtained for $D_{\rm Fi} = 0.4, 0.8, 1.2, -0.8$, and $-1.2\,$mJ/m$^2$ are displayed in Fig.\,\fref{fig:S3}{\bf b}, where a skyrmion imposed in all the layers has been used as initial state. For   $D_{\rm Fi} = 0.4\,$mJ/m$^2$ (open red up-triangles) an incomplete skyrmion is the final state even if a skyrmion spin texture is initially placed in the Fi layer. A negative DMI in the interlayer (the same sign as that of the SK layers) leads to larger skyrmion radii (compare circles to diamonds and pentagons to right-triangles for $D_{\rm Fi} = \pm 0.8$ and $\pm1.2\,$mJ/m$^2$, respectively).
 \begin{figure}
\centering
\includegraphics{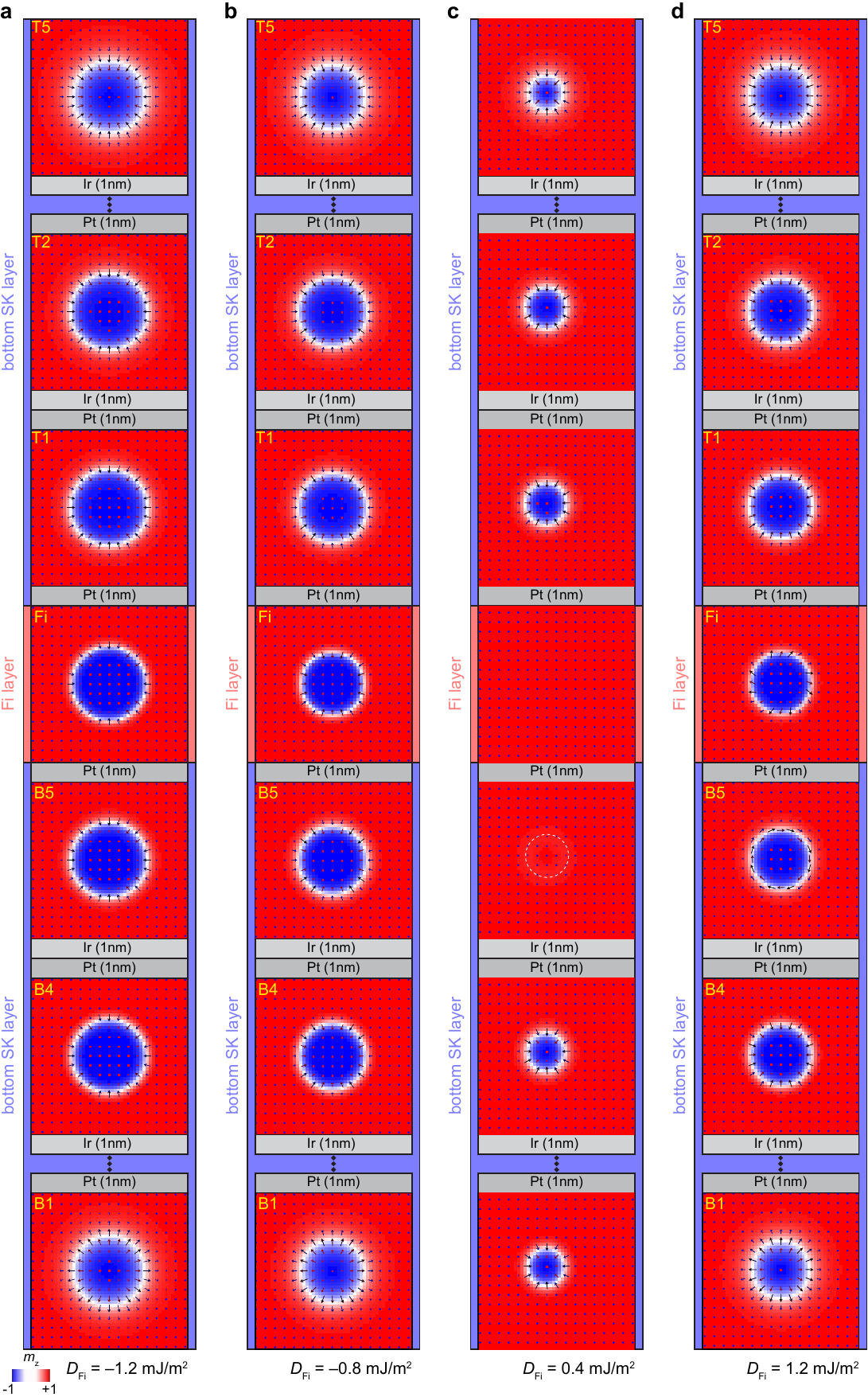}
\caption{\small {\bf a} Spatial distribution of the skyrmion magnetization as a function of the sublayer position when {\bf a} $D_{\rm Fi} = -1.2\,$mJ/m$^2$, {\bf b} $D_{\rm Fi} = -0.8\,$mJ/m$^2$, {\bf c} $D_{\rm Fi} = 0.4\,$mJ/m$^2$, and {\bf d} $D_{\rm Fi} = 1.2\,$mJ/m$^2$.}
\label{fig:S4}
\end{figure}

Figure\,\fref{fig:S4} summarizes the spatial distribution of the skyrmion magnetization (equilibrium states) corresponding to the $D_{\rm Fi}$ values reported in Fig.\,\fref{fig:S3}{\bf b}. The system is relaxed from an initial configuration where skyrmions with clockwise chirality are placed in all the layers (see Fig.\,3{\bf d} in the main text). The final equilibrium state is the result of a trade-off among negative DMI of the SK layers, different values and signs of the DMI of the Fi layer, magnetostatic interactions, and IEC between layer B5 and the first layer of the ferrimagnet.

Fig.\,\fref{fig:S4}{\bf a} shows the final state for $D_{\rm Fi} = -1.2\,$mJ/m$^2$. The tubular skyrmion has a large diameter and a clockwise chirality (N{\'e}el inward) in layers B2 to B5 and in all sublayers of the top SK layer, as determined by the negative DMI. The counter-clockwise chirality (N{\'e}el outward) of the bottom-most layer B1 permits the closure of the magnetic flux driven by the minimization of magnetostatic energy.

In Fig.\,\fref{fig:S4}{\bf b}, $D_{\rm Fi}$ is decreased to $-0.8\,$mJ/m$^2$, and the effect is only a size reduction of the tubular skyrmion, as also shown in Fig.\,\fref{fig:S3}{\bf b} (compare right-triangles to diamonds). When $-0.7\,\mbox{mJ/m}^2 < D_{\rm Fi} < 0.7\,\mbox{mJ/m}^2$, the tubular skyrmion is no longer stable and it is replaced by the incomplete skyrmion with a much smaller diameter. Fig.\,\fref{fig:S4}{\bf c} depicts an example for $D_{\rm Fi} = 0.4\,$mJ/m$^2$. Note that the chirality now is clockwise in all sublayers of the bottom SK layer. We attribute this to the suppression of the skyrmions in the Fi and in the B5 layers that lowers the {\it in-plane} component of the magnetostatic field, thus making the negative DMI the dominant field contribution.

The magnetization of the Fi layer is uniform due to the fact that its DMI is too low to sustain a skyrmion and, consequently, the magnetization of layer B5 is almost uniform as well (because of the ferromagnetic IEC considered between B5 and the first layer of the ferrimagnet). Indeed, the magnetization of layer B5 has a slight outwards canting at the location highlighted by the white dashed circle in Fig.\,\fref{fig:S4}{\bf c}. Such an outwards canting of the magnetization vectors is reminiscent of a skyrmion spin texture with a counter-clockwise chirality (N{\'e}el outward) that would occur because of the magnetostatic field from the other layers. In Fig.\,\fref{fig:S4}{\bf d} $D_{\rm Fi}$ is increased to 1.2\,mJ/m$^2$, resulting in a tubular skyrmion being stable again. However, because of the opposite signs of the DMI in the Fi and SK layers, a complex dependence of the chirality and type of spin-texture on the vertical position inside the sample occurs, as observed for the tubular skyrmion at $D_{\rm Fi} = 0.8\,$mJ/m$^2$ described in the main text. In layer B1, a skyrmion with a counter-clockwise chirality is obtained because of the minimization of the magnetostatic field. Layers B2-B4 host a skyrmion with clockwise chirality due to the negative DMI of the SK layer. The Bloch skyrmion in the B5-layer compromises between clockwise and counter-clockwise chiralities, which would derive from the negative DMI of the SK layer and from the ferromagnetic IEC with the first layer of the ferrimagnet, respectively. The skyrmion in the Fi layer is intermediate between a Bloch and a N{\'e}el with counter-clockwise chiralities. However, the skyrmion here has a chirality closer to the N{\'e}el outward due to the larger $D_{\rm Fi}$  with the respect to the case shown in the main text. The skyrmions in the top SK layer maintain a clockwise chirality as favored by the negative DMI. Our micromagnetic calculations as a function of $D_{\rm Fi}$ suggest that the ferrimagnetic layer should have a sufficiently large DMI in order to allow the coexistence of the two skyrmion phases, otherwise the tubular skyrmion would not be stable.

\pagebreak
 \section*{ \centering Supplementary note 4: Results for the coexistence of the two skyrmion phases}\vspace{-0.5cm}
 \begin{figure}[!hb]
\centering
\includegraphics[width=57.0mm]{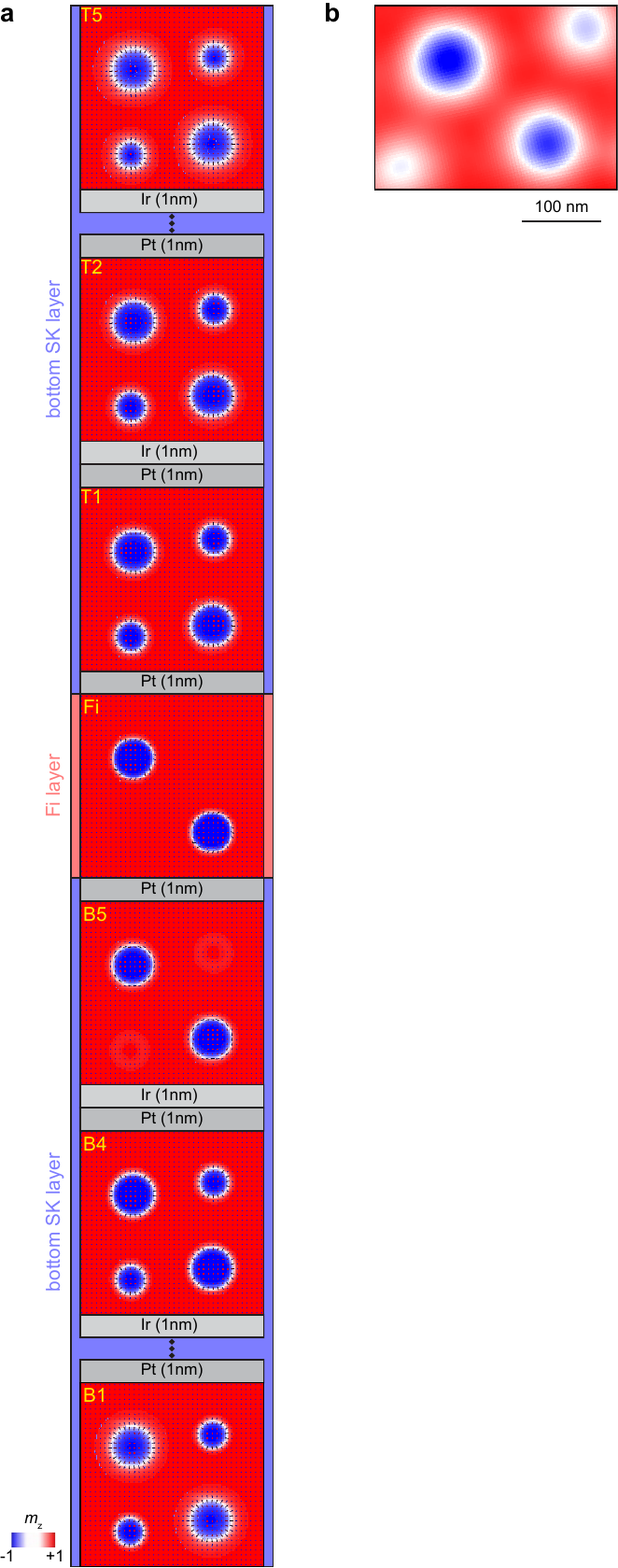}\vspace{-0.8cm}
\caption{\small {\bf a} Spatial distribution of the magnetization showing the coexistence of the tubular and incomplete skyrmions. The DMI of the SK layer is $D=-2.5\,$mJ/m$^2$, while that for the interlayer is $D_{\rm Fi}=0.8\,$mJ/m$^2$. The colors are related to the z-component of the normalized magnetization (blue negative, red positive). {\bf b} MFM subimage cut from Fig.\,1{\bf e} rotated to resemble the configuration of tubular and incomplete skyrmions of {\bf a}.}
\label{fig:S5}
\end{figure}

\pagebreak
